\documentclass[aps,prl,twocolumn,showpacs,superscriptaddress]{revtex4-1}  
\bibpunct{[}{]}{,}{n}{}{}
\usepackage{graphicx}
\usepackage{dcolumn}
\usepackage{bbm}
\usepackage{latexsym}
\usepackage{lipsum}
\usepackage{amssymb}
\usepackage{amsfonts}
\usepackage{amsmath}
\usepackage{fancybox}
\usepackage{mathtools}
\usepackage[binary,squaren]{SIunits}
\usepackage{color}
\usepackage{colordvi}%
\providecommand{\U}[1]{\protect\rule{.1in}{.1in}}

\def\be{\begin{equation}}
\def\ee{\end{equation}}
\def\ber{\begin{eqnarray}}
\def\eer{\end{eqnarray}}

\begin{document}

\setlength{\abovedisplayskip}{3.5pt}
\setlength{\belowdisplayskip}{3.5pt}

\preprint{APS/123-QED}

\title{Magnetoresistance driven by the magnetic Berezinskii-Kosterlitz-Thouless transition}

\author{B. Flebus}
\affiliation{Department of Physics, Boston College, 140 Commonwealth Avenue Chestnut Hill, MA 02467}

\date{\today}

\begin{abstract}

While the Berezinskii-Kosterlitz-Thouless transition (BKT) has been under intense scrutiny for decades,  unambiguous experimental signatures in magnetic systems remain elusive. Here, we investigate the interplay between electronic and magnetic  degrees of freedom near the BKT transition. Focusing on a metal with easy-plane ferromagnetic order, we establish a framework  that accounts both for the coupling between the charge current  and the flow of topological magnetic defects and for electron scattering on their inhomogeneous spin texture. We show that electron scattering  is responsible for  a temperature-dependent magnetoresistance effect scaling as the density of the topological defects, which is expected to increase dramatically above  the BKT transition temperature. Our findings call for further experimental investigations.

\end{abstract}

\maketitle

It has been decades since Berezinskii, Kosterlitz and Thouless  predicted  a continuous topological transition  in the two-dimensional (2$d$) XY model~\cite{paper1,paper2,paper3}. The BKT transition represents one of 
the first examples of a topological transition beyond the Landau-Ginzburg paradigm, i.e., not driven by symmetry breaking, but associated with the binding and unbinding of topological defects. The low-temperature phase of the XY model is stabilized by the formation of bound vortex–antivortex pairs. At  the 
transition temperature $T_{\text{BKT}} \neq 0$,  the free energy is instead minimized by  pairs unbinding  and defects profileration. 
Above the transition temperature, the density of topological defects increases exponentially with increasing temperature, until their correlation length approaches the lattice constant and the system enters in a highly disordered phase~\cite{paper3}. 

While historically the BKT transition was proposed in a XY spin model, experimental signatures  were first observed in superfluid helium films~\cite{superfluidfilm1,superfluidfilm2}, 2$d$ superconductors~\cite{superconductingfilm1,superconductingfilm2} and arrays of Josephson junctions~\cite{array}.  More recently, evidence of the BKT transition has been reported in trapped Bose gases~\cite{Bosegas1,Bosegas2,Bosegas3}. Despite  intense experimental efforts~\cite{BKT0,BKT1,BKT2,BKT3,BKT4, BKT41,BKT5,BKT6}, nevertheless,  unambiguous evidence of the BKT transition in a magnetic system is still lacking. Several theoretical studies have addressed the  properties of quasi-2$d$ layered magnetic compounds with weak interplane interactions~\cite{3d1,3d2,3d3,3d4,montecarlo} and of 2$d$ van der Waals magnets near the BKT transition temperature~\cite{vdW,vdW1}. Signatures of the BKT transition in the critical behavior of  spin-spin correlation functions and spin currents have been extensively investigated. 
On the other hand, it remains still relatively  unexplored how evidence of a BKT transition could be uncovered in the properties of electronic or phononic degrees of freedom interacting with the  topological magnetic  defects.

Recently a colossal magnetoresistance (CMR) was observed in Eu$\text{Cd}_{2}\text{P}_{2}$~\cite{Wang}, a compound comprised of ferromagnetic layers weakly  coupled via planes intercalated with non-magnetic atoms. 
While the experimental results have not been yet fully understood, it is clear that Eu$\text{Cd}_{2}\text{P}_{2}$ does not fit the mixed valence CMR paradigm. Instead,  experimental evidence points at an electrical resistivity driven by 2$d$ magnetic fluctuations that grow dramatically far above the 3$d$ magnetic ordering temperature  and are strongly suppressed by a magnetic field. 

Inspired by these results, here we investigate the coupling between an electric current and magnetic topological defects  near the  BKT transition temperature. We consider a 2$d$ anisotropic easy-plane Heisenberg  model, which falls into the universality class of 2$d$ XY spin systems~\cite{3d3}. 
\begin{figure}[b!]
\includegraphics[width=1\linewidth]{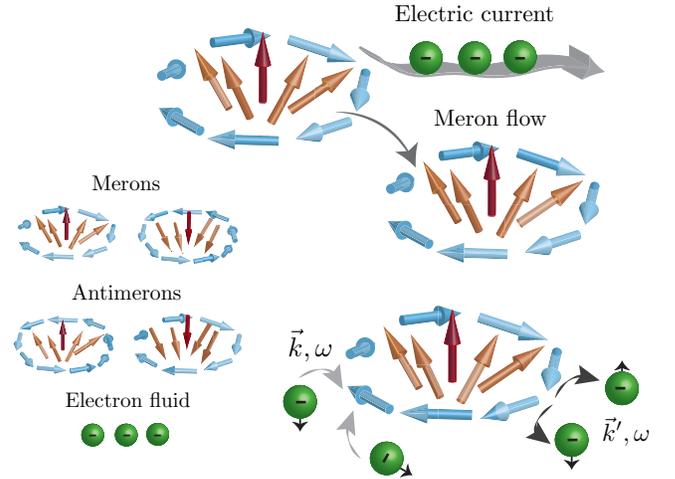}
\caption{Above the BKT transition temperature, a  metallic system with  easy-plane ferromagnetic order can  host an electron fluid and magnetic topological defects, i.e., merons and antimerons.   The electric current interacts with the flow of topological defects, and electrons scatter elastically on the inhomogeneous magnetization textures of merons and antimerons. In this work, we address the contribution of these processes to the electrical resistivity.}
\label{Fig1}
\end{figure} This model can host vortex-like spin configurations, characterized by a topological charge dubbed as the meron number $Q$~\cite{meron1}. We focus on a temperature regime above the transition temperature, in which merons ($Q>0)$ and anti-merons ($Q<0)$ can be treated as free particles. Using Onsager reciprocity and the collective coordinate approach, we derive an effective two-fluid model, comprised of an electron fluid and a fluid of magnetic topological defects.

 As depicted in Fig.~\ref{Fig1}, we investigate the corrections to the electric resistivity arising from two distinct processes: the interaction between the electronic flow and  the currents of topological defects; and  electron scattering on the inhomogeneous spin texture of the defects. Using Boltzmann theory for diffusive electronic transport, we show that scattering dominates the longitudinal electric resistivity, leading to a correction proportional to the density of topological defects, which grows exponentially with temperature. In the presence of an in-plane magnetic field, the BKT transition and, consequently, the proliferation of topological defects are suppressed~\cite{supp}.
Thus, our results suggest that a magnetic BKT transition in a metallic system can lead to a colossal temperature-dependent magnetoresistance effect.

 \textit{Model.}  We consider a metallic film with easy-plane ferromagnetic order. The film is set at a temperature $T \gtrsim T_{\text{BKT}}$ and subjected to an external electric field $\vec{E}$. The electronic system is treated as a Fermi gas with electron density $n_{e}$ and current flow $\vec{j}_{e}= n_{e} \vec{v}_{e}$, where $\vec{v}_{e}$ is the electron drift velocity.
 The magnetic system is described by an easy-plane Heisenberg Hamiltonian, which, in the continuum limit, can be written as~\cite{3d3}
 \begin{align}
\mathcal{H}=\frac{J}{2}\int d^{2} \vec{r} \bigg[& \left( 1-\frac{\lambda}{2} \cos^2\theta \right) \left( \nabla \theta \right)^2+\sin^2\theta \left( \nabla \phi\right)^2 \nonumber \\
&+\lambda \cos^2\theta\bigg]\,.
\label{87}
 \end{align}
Here, $\theta=\theta(\vec{r})$ and $\phi =\phi(\vec{r})$ are, respectively, the polar and azimuthal angle of the ferromagnetic order parameter $\mathbf{n}(\vec{r})$. $J$ parametrizes the strength of the ferromagnetic exchange interaction and $\lambda$  is the anisotropic parameter, with $0 < \lambda \leq 2$.
Focusing on a square lattice, the model~(\ref{87}) admits  as stable meron solutions for $\lambda \lesssim 0.6 $~\cite{wysin1,wysin2}. 
Far away from the meron core, $\mathbf{n}(\mathbf{r})$ lies in the easy plane and forms a planar vortex  characterized by a winding number $\nu=\pm 1$. In the core region  $\mathbf{n}(\mathbf{r})$  smoothly rotates either up or down out of the easy plane: the direction along which the core magnetization points defines its polarity  $p=\pm 1$. The meron  number  $Q= p \nu /2$ is equivalent to half-skyrmion number~\cite{belavin}, i.e., 
\begin{align}
Q=\frac{1}{8\pi} \int d^{2}\mathbf{r} \; \mathbf{n} \cdot \left( \partial_{x}\mathbf{n} \times \partial_{y}\mathbf{n} \right)\,.
\end{align}
Below the BKT transition,  topological defects  carrying opposite charge are bound in meron-antimeron pairs. 
Above the BKT transition temperature $T_{\text{BKT}}$, the correlation length $\xi$ of the topological defects, divergent at $T_{\text{BKT}}$, decays rapidly as function of temperature. Adopting as a guide the results of the XY model, which corresponds to Eq.~(\ref{87}) for $\lambda=1$, from renormalization group analysis one finds~\cite{paper3} 
\begin{align}
\xi(T) = a_{0} e^{b/ \sqrt{\tau}}\,, \; \; \; \; \; \tau=\left(T-T_{\text{BKT}}\right)/T_{\text{BKT}}\,,
\label{correlength}
\end{align} 
where $a_{0}$ is of the order of the lattice constant $a$ and $b \approx 1.5$ is a non-universal constant. The correlation length (\ref{correlength}) can be interpreted as half of the mean separation between free merons~\cite{corr1,corr2}. Thus, we can estimate the  density $n$ of free topological defects as
\begin{align}
n \equiv n(T) \approx  \xi(T) ^{-2}\,.
\label{105}
\end{align}

In the regime of validity of Eq.~(\ref{correlength}), i.e., for $T \gtrsim T_{\text{BKT}}$,  merons and antimerons can be treated  as (approximately) free particles. 
Using the collective coordinate approach~\cite{thiele}, we model each defect  as a rigid particle moving at the drift velocity $\vec{v}$ of its center of mass. For a density $n_{Q}$ of topological defects with charge $Q$, we can introduce the  current $\vec{j}_{Q}=n_{Q}\vec{v}$, whose dynamics obeys~\cite{thiele}
\begin{align}
\left[ \hat{\mathcal{D}} + \hat{\mathcal{G}}\right]  \cdot  \vec{j}_{Q}=\vec{F} n_{Q}\,.
\label{98}
\end{align}
Here,  $s$ is the sheet spin density,  $\vec{F}$ is the force acting on the topological defects, $\mathcal{G}_{ij}=4\pi s Q \epsilon_{ji}$ is the gyromagnetic tensor and  $\hat{\mathcal{D}}$ is the viscosity tensor, with components
\begin{align}
\mathcal{D}_{ij}=\alpha s \int d^{2} \vec{r} \; \partial_{i} \mathbf{n} \cdot \partial_{j} \mathbf{n} \,,
\end{align}
where $\alpha$ is the (dimensionless) Gilbert damping and $s$ the (sheet) spin density. For an axially-symmetric spin texture,  we can set $\mathcal{D}=\mathcal{D}_{xx(yy)}$ and $\mathcal{D}_{ij} = 0$, for $i\neq j$~\cite{disstens}.

An electric current exerts an adiabatic and a non-adiabatic spin transfer torque on the spin texture of the topological defects~\cite{Slo, Berger, Tatara, Jones, STT1, STT}. Within the collective coordinate approach,  the spin-transfer torques can be accounted for by rewriting Eq.~(\ref{98}) as~\cite{STT}
\begin{align}
 \hat{\mathcal{D}} \cdot \left( \vec{j}_{Q}-\frac{\beta}{\alpha} \frac{n}{n_{e}} \mathcal{P} \vec{j}_{e}\right)+ \hat{\mathcal{G}}\cdot \left( \vec{j}_{Q}-\frac{n}{n_{e}} \mathcal{P} \vec{j}_{e} \right)=\vec{F} n_{Q}\,,
 \label{STT}
\end{align}
where $\mathcal{P}$ is a dimensionless phenomenological parameter, and the (dimensionless) constant $\beta$  describes the coupling between current and local magnetization owing to non-adiabatic effects.

\begin{figure}[t!]
\includegraphics[width=1\linewidth]{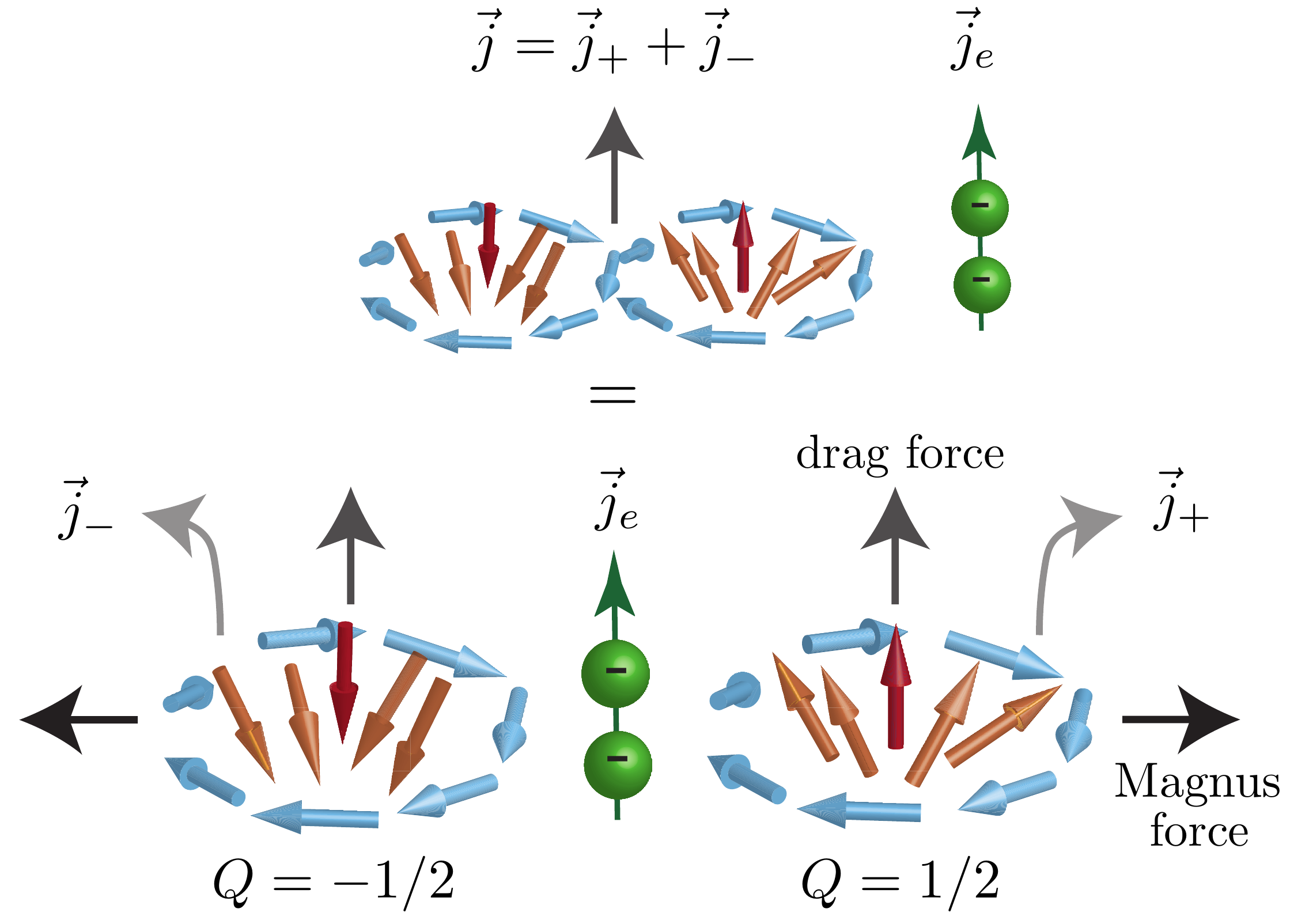}
\caption{ Interplay between the electric current, $\vec{j}_{e}$, and the meron, $\vec{j}_{+}$ and antimeron, $\vec{j}_{-}$, currents. The viscous drag force drags the meron and antimeron currents along the electron flow. The Magnus force  results into the transverse motion of topological defects with respect to the driving current.  If the meron and antimeron density are equal, the meron and antimeron Hall currents cancel out. The system can be then described in terms of  a total current of topological defects $\vec{j}=\vec{j}_{+}+\vec{j}_{-}$ interacting with the electric current $\vec{j}_{e}$ via a drag force.}
\label{Fig2}
\end{figure}

The magnetic phase above the BKT transition can be generally described in terms of a meron fluid  with density $(n+ \delta n)/2$ and current $\vec{j}_{+}$, and an antimeron  fluid  with    density $(n- \delta n)/2$ and current $\vec{j}_{-}$. However, due to the symmetry of the Hamiltonian~(\ref{87}) under the reflection $\mathbf{n}(\mathbf{r})\rightarrow - \mathbf{n}(\mathbf{r})$, merons and antimerons have the same energy. Thus, they proliferate with equal probability, i.e., $\delta n=0$.
Equation~(\ref{STT}) shows that  the topological defects are subjected to a drag force $\propto \mathcal{D} \cdot \vec{j}_{e}$ in the direction of the driving current, and to a topological Magnus force $\propto \mathcal{G} \cdot \vec{j}_{e}$  that results in their transverse (gyrotropic) motion with respect to the driving current. The direction of the transverse (Hall) current of topological defects is dictated by the sign of the charge $Q$, as shown in Fig.~\ref{Fig2}.
For an equal density $n/2$ of merons and antimerons, the  meron and antimeron Hall currents cancel out, i.e.,
\begin{align}
\left(\vec{j}_{+} \right)_{\bot \vec{j}_{e}}=- \left(\vec{j}_{-} \right)_{\bot \vec{j}_{e}}\,.
\label{177}
\end{align}  
When Eq.~(\ref{177}) holds, we can describe the magnetic dynamics in terms of  a total current of topological defects $\vec{j}=\vec{j}_{+}+\vec{j}_{-}=n\vec{v}$ that  is coupled to the electric current only via a viscous drag force. The Onsager matrix reads as~\cite{onsager}
\begin{align}
\begin{pmatrix} \vec{F}_{e} \\ \vec{F}  \end{pmatrix}=\begin{bmatrix}
e^2 \hat{\rho}_{e} & -(\beta \mathcal{P}/\alpha n_{e})\hat{\mathcal{D}}  \\ -(\beta \mathcal{P}/\alpha n_{e})\hat{\mathcal{D}} & (1/n) \hat{\mathcal{D}}
\end{bmatrix} \begin{pmatrix} \vec{j}_{e} \\ \vec{j}  \end{pmatrix}\,,
\label{317}
\end{align}
where  $\vec{F}_{e}=-e\vec{E}$ is the force exerted by the electric field on a charge $-e$ (with $e>0$) and $\hat{\rho}_{e}$ is the electrical resistivity tensor, which can be written as
\begin{align}
\hat{\rho}_{e}=\left( \rho + \rho_{v} \right) \mathbb{I}\,,
\label{resistivity1}
\end{align}
where $\mathbb{I}$ is the 2$\times$2  identity matrix. Here, $\rho$ is the electrical resistivity independent of the magnetic degrees of freedom, while the resistivity $\rho_{v}$ is due to  electron scattering on  topological magnetic defects. 
 By solving Eq.~(\ref{317}) while setting $\vec{F}=0$, we find
\begin{align}
\hat{\rho}_{e}=\left[ \rho + \rho_{v} - \frac{n s}{n^2_{e}}  \frac{\beta^2  \mathcal{P}^2  \mathcal{D}}{ e^2 \alpha}  \right] \mathbb{I}\,.
\label{137}
\end{align}
Equation~(\ref{137}) shows that the drag exerted by the flow of topological defects leads to a reduction of the electrical resistivity~(\ref{105}). This correction, however, might be negligible for good metals even for an exponentially increasing density of topological defects,   as it scales as $\propto sn/n^2_{e}$.

\textit{Scattering}. We proceed to address the contribution to the electron resistivity, $\rho_{v}$, due to  elastic scattering on the inhomogeneous magnetization profile of the topological defects. The electron mean free path $\ell_{\text{mfp}}$ is taken to be much larger than the topological defect correlation length $\xi$~(\ref{correlength}), i.e., $\ell_{\text{mfp}} \gg \xi$. Under this assumption, we can treat the electron  dynamics  as diffusive~\cite{RMF}.

A solution of  Eq.~(\ref{87}) can not be found in a simple analytical form;  however,  we can use a parametrization of the spin texture that it is compatible with its asymptotic form. For a topological defect with polarity $p$ and winding number $\nu$, we write
\begin{align}
\mathbf{n}(\vec{r})=\begin{pmatrix} \sqrt{1-e^{-\frac{2r}{a_{c}}}}\cos \nu \phi, \sqrt{1-e^{-\frac{2r}{a_{c}}}} \sin \nu \phi,  p e^{-\frac{r}{a_{c}}} \end{pmatrix}\,.
\end{align} 
where $r=|\vec{r}|$ and $a_{c}=a/\sqrt{\lambda}$ can be interpreted as the vortex core radius.
The electron spin $\hat{\mathbf{s}}$ interacts with the inhomogeneous spin texture $\mathbf{n}(\vec{r})$ via an exchange interaction, whose strength is parametrized by $J_{sd}$ as
\begin{align}
V(\vec{r})=-  J_{sd}  s  \mathbf{n}(\vec{r})  \cdot \hat{\mathbf{s}}\,.
\label{potential}
\end{align}
 In the leading Born approximation, the scattering matrix element associated with Eq.~(\ref{potential}) reads as
\begin{align}
F_{\alpha \beta}(\vec{k}-\vec{k}')=& \int d^{2}\vec{r} \; e^{-i(\vec{k}'-\vec{k})\cdot \vec{r}} \langle \vec{k}', \alpha | V(\vec{r}) | \vec{k}, \beta \rangle\,,
\label{144}
\end{align}
where electron initial state $|\vec{k},\beta\rangle$ is a plane wave with momentum $\vec{k}$ and spin $\beta$, while the electron final state $|\vec{k}',\alpha \rangle$ is a plane wave with momentum $\vec{k}'$ and spin $\alpha$, with $|\vec{k}|,|\vec{k}'|=k$.
 For an equal density of defects with polarity $p=\pm 1$,  we have $F_{\uparrow \downarrow}=F_{\downarrow \uparrow}$. To obtain an estimate of the order of magnitude of $\rho_{v}$,  here we focus  for simplicity on spin-conserving scattering processes. By plugging Eq.~(\ref{potential}) into Eq.~(\ref{144}),  the spin-conserving scattering matrix elements can be found as
\begin{align}
F_{\uparrow \uparrow (\downarrow \downarrow)}(\vec{k}-\vec{k}')=\mp  \frac{  \pi  p J_{sd} s a^2_{c} }{\left(1+4 k^2 a^2_{c} \sin^2\frac{\phi_{sc}}{2}\right)^{3/2}}\,,
\label{eq17}
\end{align}
where $\phi_{sc}$ is the scattering angle.
For a steady homogeneous state, the  Boltzmann equation for the electron distribution function $f_{\vec{k}}=f^{(0)}_{\vec{k}}+f^{(1)}_{\vec{k}}$ 
reads as 
\begin{align}
\left( \frac{\partial f}{\partial t} \right)_{\text{coll}}
=-\frac{\partial f^{(0)}}{\partial \epsilon_{\vec{k}}}  \vec{v}_{\vec{k}} \cdot e \vec{E}\,.
\label{155}
\end{align}
Here, $f^{(0)}_{\vec{k}}$ and $f^{(1)}_{\vec{k}}$ are, respectively, the equilibrium and non-equilibrium component of the Fermi distribution of a Fermi gas with isotropic dispersion $\epsilon_{k}=\hbar^2 |\vec{k}|^2/(2m^{*})$ and velocity $\vec{v}_{\vec{k}}=(1/\hbar)\partial_{\vec{k}} \epsilon_{\vec{k}}$, where $m^{*}$ is the electron effective mass.  Using Fermi Golden's rule, we can rewrite the collision integral associated with Eq.~(\ref{eq17}) as
\begin{align}
\left( \frac{\partial f_{\vec{k}}}{\partial t} \right)_{\text{coll}}&=
\frac{2\pi n}{\hbar} \int \frac{d^2\vec{k}'}{(2\pi)^2} |F_{\uparrow \uparrow}(\vec{k}-\vec{k}')|^2 \delta \left( \epsilon_{\vec{k}}-\epsilon_{\vec{k}'} \right) \nonumber 
\\ &\times  \left( f^{(1)}_{\vec{k}'}- f^{(1)}_{\vec{k}}\right)\,.
\label{163}
\end{align}
Plugging into Eq.~(\ref{163}) the following ansatz for the non-equilibrium component of the distribution function:
\begin{align}
f^{(1)}_{\vec{k}}=\tau_{v}(\epsilon_{\vec{k}}) e \vec{E} \cdot \vec{v}_{\vec{k}} \frac{\partial f^{0}}{\partial \epsilon|_{\epsilon_{\vec{k}}}}\,,
\end{align}
we find the electron relaxation time due to scattering on topological defects as 
\begin{align}
\frac{1}{\tau_{v}(\epsilon_{k})} = n m^{*}
\left(\frac{  \pi p a^2_{c}J_{sd} s}{ \hbar} \right)^2    \frac{1+(a_{c}k)^2}{[1+(4a_{c}k)^2]^{\frac{5}{2}}}\,.
\end{align}
Electron-impurity  and  electron-phonon scattering contribute as well to the electron relaxation time.   Here, we focus for simplicity on  elastic scattering on nonmagnetic impurities.
We assume no interference between electron scattering on magnetic and nonmagnetic impurities.  For $T \ll T_{F}$, where $T_{F}$ is the Fermi temperature,  the longitudinal resistivity $\rho_{e}$~(\ref{resistivity1}) can be  written as
\begin{align}
\rho_{e}=\frac{ m^*}{ e^2 n_{e} }\left[ \frac{1}{\tau_{i}(\epsilon_{F})}+ \frac{1}{\tau_{v}(\epsilon_{F})} \right]\,,
\label{670}
\end{align}
where $\epsilon_{F}$ is the Fermi energy.  Here, $\tau_{i}(\epsilon_{F}) \propto v_{F}/n_{i} \Sigma$ is the relaxation time due to scattering on non-magnetic impurities, where $v_{F}$ is the Fermi velocity, $\Sigma$ the hard-sphere impurity scattering cross section and $n_{i}$ the impurity density. 
Equations~(\ref{correlength}),~(\ref{105}) and~(\ref{670}) show that, for $T \gtrsim T_{\text{BKT}}$, the temperature dependence of the resistivity is dominated by the exponential growth of topological defects, i.e., $\rho_{e}(T) \propto n(T)$. In the presence of an in-plane magnetic field $B$, the Hamiltonian~(\ref{87}) admits a single ground-state stable solution for $\phi=\text{const}$, i.e., a uniform planar spin texture. Vortex-like excitations can appear as the temperature is raised~\cite{Gouveat}, but the Kosterlitz-Thouless transition is suppressed. Thus, our results point at a temperature-dependent magnetoresistance that scales as $\rho(B=0,T)-\rho(B,T) \propto n(T)$.


\textit{Discussion and outlook.} In this work, we constructed a  theory that describes the interactions between electric current and magnetic topological defects in a metal with easy-plane ferromagnetic order above the BKT transition. While our framework is suited to describe the interplay between a Fermi gas and topological defects obeying a Landau-Lifshitz-type dynamics in a $U$(1)-symmetric spin system, it can be easily generalized. 

We investigated the correction to the electron resistivity due to the coupling between the electron and meron flow and due to electron scattering on the inhomogeneous spin textures of topological defects, finding that the latter dominates. We show that the  magnetoresistance scales as the density of topological defects, i.e., it grows exponentially  above the transition temperature.  Our results, thus, suggest that  temperature-dependent colossal magnetoresistances in  quasi-$2d$ or $2d$ magnetic metals  that do not fit the mixed valence CMR paradigm~\cite{Wang} should be further investigated as a possible signature of a magnetic BKT transition. 

A systematic study of the role of spin waves and of the magnon drag on the electric current~\cite{drag} in this regime are called upon. The role of a weak out-of-plane magnetic field, which lifts the degeneracy of the meron-antimeron solutions, will be addressed in future work.
\\ \\
The author is grateful to F. Tafti and Y. Tserkovnyak for inspiring discussions.

\end{document}